\documentclass[a4paper,11pt]{article}
\pdfoutput=1 
\usepackage{graphicx}
\usepackage{jcappub} 

\usepackage[T1]{fontenc} 

\title{Testing dark energy models with $H(z)$ data}

\author[a]{Jing-Zhao Qi,}
\author[b]{Ming-Jian Zhang,}
\author[a,1]{Wen-Biao Liu\note{Corresponding author.}}

\affiliation[a]{Department of Physics, Institute of Theoretical Physics, Beijing Normal University, Beijing, 100875, China}
\affiliation[b]{Key Laboratory of Particle Astrophysics, Institute of High Energy Physics,
Chinese Academy of Science, P. O. Box 918-3, Beijing 100049, China}

\emailAdd{wbliu@bnu.edu.cn}

\abstract{
$Om(z)$ is a diagnostic approach to distinguish dark energy models. However, there are few articles to discuss what is the distinguishing criterion. In this paper, firstly we smooth the latest observational $H(z)$ data using a model-independent method -- Gaussian processes, and then reconstruct the $Om(z)$ and its fist order derivative $\mathcal{L}^{(1)}_m$. Such reconstructions not only could be the distinguishing criteria, but also could be used to estimate the authenticity of models. We choose some popular models to study, such as $\Lambda$CDM, generalized Chaplygin gas (GCG) model, Chevallier-Polarski-Linder (CPL) parametrization and Jassal-Bagla-Padmanabhan (JBP) parametrization. We plot the trajectories of $Om(z)$ and $\mathcal{L}^{(1)}_m$ with $1 \sigma$ confidence level of these models, and compare them to the reconstruction from $H(z)$ data set. The result indicates that the $H(z)$ data does not favor the CPL and JBP models at $1 \sigma$ confidence level. Strangely, in high redshift range, the reconstructed $\mathcal{L}^{(1)}_m$ has a tendency of deviation from theoretical value, which demonstrates these models are disagreeable with high redshift $H(z)$ data. This result supports the conclusions of Sahni et al. \citep{sahni2014model} and Ding et al. \citep{ding2015there} that the $\Lambda$CDM may not be the best description of our universe.}

\begin{document}
\maketitle
\flushbottom

\section{Introduction}  \label{introduction}

The late time cosmic acceleration has been supported by many independent cosmological observations, including the type Ia supernovae (SNIa) \citep{riess1998supernova}, large scale structure \citep{tegmark2004cosmological}, cosmic microwave background (CMB) anisotropy \citep{spergel2003wmap} etc. An additional component, dubbed as dark energy, has been proposed to explain this phenomenon.
Dark energy with an equation of state (EoS) $w$---ratio of its pressure and energy density is believed to be an impetus of the cosmic acceleration. According to the EoS, many candidates can be a possibility of the mysterious dark energy. The cosmological constant model namely, $\Lambda$CDM with $w=-1$ is the most robust model. However, it suffers the notable fine-tuning problem \citep{weinberg1989cosmological,weinberg2000cosmological} and coincidence problem \citep{1999PhRvL..82..896Z}. A dark energy without a constant vacuum energy naturally becomes a widespread speculation. Hence then, a number of dynamical dark energy models have been proposed, such as quintessence \citep{Caldwell:2005tm,Zlatev:1998tr,Tsujikawa:2013fta}, K-essence \citep{chiba2000kinetically,armendariz2000dynamical}, phantom \citep{kahya2007quantum,onemli2004quantum,singh2003cosmological}, Chaplygin gas \citep{bento2002generalized,kamenshchik2001alternative}, and so on. On the other hand, plentiful parameterized EoS also have been widely employed to analyse the behavior of dark energy \citep{riess2004type,barboza2009generalized,zhang2015exploring,maor2001limitations,
chevallier2001accelerating,linder2003exploring,Jassal:2004ej,Wei:2013jya}.

In the grand dark energy family, most of them are consistent with the observational data. The burden, therefore falls on the question of which model is realistic, so can we ascertain the unique truth? This may be philosophical. But we should try to distinguish the increasing numbers of dark energy models, and try to exclude some of them. Fortunately, there have been some geometrical diagnostics, such as Statefinder \citep{Sahni:2002fz}, $Om$ diagnostic \citep{sahni2008two}, Statefinder hierarchy \citep{Arabsalmani:2011fz}, which could be used to distinguish dark energy models. The related research can refer to Refs. \citep{Sahni:2002fz,shojai2009statefinder,zhang2005statefinder,bao2007statefinder,chang2007statefinder,feng2008statefinder,sahni2008two,Arabsalmani:2011fz,Li:2014mua,qi2015several}. The principle of such diagnostics is that different models will show different evolutionary trajectories in defined parameters plane. If the distances between such trajectories are far enough, it can be concluded that these models could be discriminated. But, there are few articles to discuss how far away is the distinguishing criterion. Theoretically, this criterion should depend on the observational precision. Besides, it should be independent of cosmological model. However, the parameters of these diagnostics --- Statefinder's $\{r,s\}$, $Om$'s $Om(z)$ and Statefinder hierarchy $a^{(n)}/aH^n$, $n\geq3$ --- are not observable quantity. They cannot directly compare with observational data. However, we could reconstruct these parameters of diagnostics from existing observational data. Due to the observational data have errors, the reconstructions will be with error ranges. Such reconstructions of diagnostic parameters could be the distinguishing criterion. If the distances between trajectories of models are greater than the error range of reconstruction, these models could be discriminated. Meanwhile, once the trajectories of models are beyond the error range of reconstructions, we could doubt the authenticity of these models.
Therefore, such reconstructions not only could be the distinguishing criteria, but also could be used to estimate the authenticity of models. In order to obtain the reconstructions, we need to smooth model-independently the existing data, and to estimate the derivatives. Fortunately, Gaussian processes (GP) can meet this challenges.

In this paper, we focus on $Om(z)$ diagnostic and its first derivative $\mathcal{L}^{(1)}_m$. In comparison to Statefinder constructed by the third and higher order derivatives of the scale factor $a(t)$, $Om(z)$ just use the first order derivative of $a(t)$. Thus, $Om(z)$ is a preferred choice to apply to observational data. In addition, $Om(z)$ could provide a null test on the $\Lambda$CDM. Namely, if dark energy is the cosmological constant, the value of $Om(z)$ is constant. A positive and negative slope of $Om(z)$ represent phantom and quintessence models respectively. The first derivative of $Om(z)$, $\mathcal{L}^{(1)}_m$, provides more effective test that measures deviations from zero easier than from a constant, i.e. $\mathcal{L}^{(1)}_m=0$ for $\Lambda$CDM, $\mathcal{L}^{(1)}_m>0$ for phantom and $\mathcal{L}^{(1)}_m<0$ for quintessence.
We will reconstruct the $Om(z)$ and $\mathcal{L}^{(1)}_m$ from the observational $H(z)$ data. In Ref. \citep{seikel2012using}, the $Om(z)$ and $\mathcal{L}^{(1)}_m$ have been reconstructed only on the consistency tests of the $\Lambda$CDM model. For a further analysis, we will perform consistency test on more dark energy models including $\Lambda$CDM, the generalized Chaplygin (GCG), Chevallier-Polarski-Linder (CPL) parametrization model and Jassal-Bagla-Padmanabhan (JBP) parametrization model. We intend to discriminate these models and test their authenticity. Because such reconstructions are completely given by the observed data and are model-independent, the model consists with them better means that it is more realistic. In addition, the $H(z)$ data used in Ref. \citep{seikel2012using} is not latest. We will reconstruct the $Om(z)$ and $\mathcal{L}^{(1)}_m$ from the latest $H(z)$ data to compare with above dark energy models.

This paper is organized as follows. In Sec. \ref{2}, we briefly revisit $Om$ diagnostic and its first-order derivative $\mathcal{L}^{(1)}_m$ as another effective discriminating quantity, and show their reconstruction by $H(z)$ data using Gaussian processes. In Sec. \ref{DE}, some dark energy models including $\Lambda$CDM, GCG, CPL and JBP are introduced. In Sec. \ref{RD}, $Om(z)$ and $\mathcal{L}^{(1)}_m$ of these models are compared with their reconstructions from $H(z)$ data. According to comparisons, some discussions are given. Finally, the conclusion is presented in Sec. \ref{Con}.

\section{Theoretical method} \label{2}

In the Friedmann-Robertson-Walker spacetime, we have the general Friedmann equation
\begin{eqnarray}
h^2(z)&\equiv & \frac{H^2(z)}{H^2_0}  \nonumber \\
&=& \Omega_m(1+z)^3+\Omega_K(1+z)^2+(1-\Omega_m-\Omega_K) \nonumber \\
&&\times \exp \left(3\int_0^z \frac{1+w(z')}{1+z'}dz' \right). \label{fried}
\end{eqnarray}
Using Eq. (\ref{fried}), we can define $Om$ diagnostic function over the redshift $z$ \citep{sahni2008two,shafieloo2010model}
\begin{eqnarray}
Om(z)\equiv \frac{h^2(z)-1}{(1+z)^3-1} .  \label{om}
\end{eqnarray}
For the $\Lambda$CDM, the value of $Om(z)$ is a constant independent of the redshift. Therefore, if $Om(z)$ is variable, it possibly leads to an alternative dark energy or modified gravity model. We can reconstruct $Om(z)$ from observed $h(z)$ data to test a series of dark energy models. As mentioned above, we could define another effective quantity $\mathcal{L}^{(1)}_m$ \citep{seikel2012using}

\begin{eqnarray}
\mathcal{L}^{(1)}_m &\equiv & \frac{1}{(1+z)^6}\frac{dOm(z)}{dz} \nonumber \\
&=&\frac{3(1+z)^2(1-h^2)+2z(3+3z+z^2)hh'}{(1+z)^6}. \label{L1}
\end{eqnarray}
$\mathcal{L}^{(1)}_m=0$ implies the $\Lambda$CDM. To obtain this quantity, $h'(z)$ needs to be constructed from the data. It is crucial to employ a model-independent method to reconstruct $h(z)$ and its derivative $h'(z)$. There is a suitable approach, the so-called Gaussian processes can accomplish this task. Here we use the publicly available code GaPP (Gaussian processes in Python). Its algorithm could be found in Ref.  \citep{seikel2012reconstruction} and on the Gaussian Process webpage \citep{gapp}. This GP code has been widely used in many works \citep{seikel2012reconstruction,yahya2014null,cai2015reconstructing,seikel2013optimising,cai2015null,seikel2012using}.

\begin{figure}
\centering
\includegraphics[width=16cm,height=6cm]{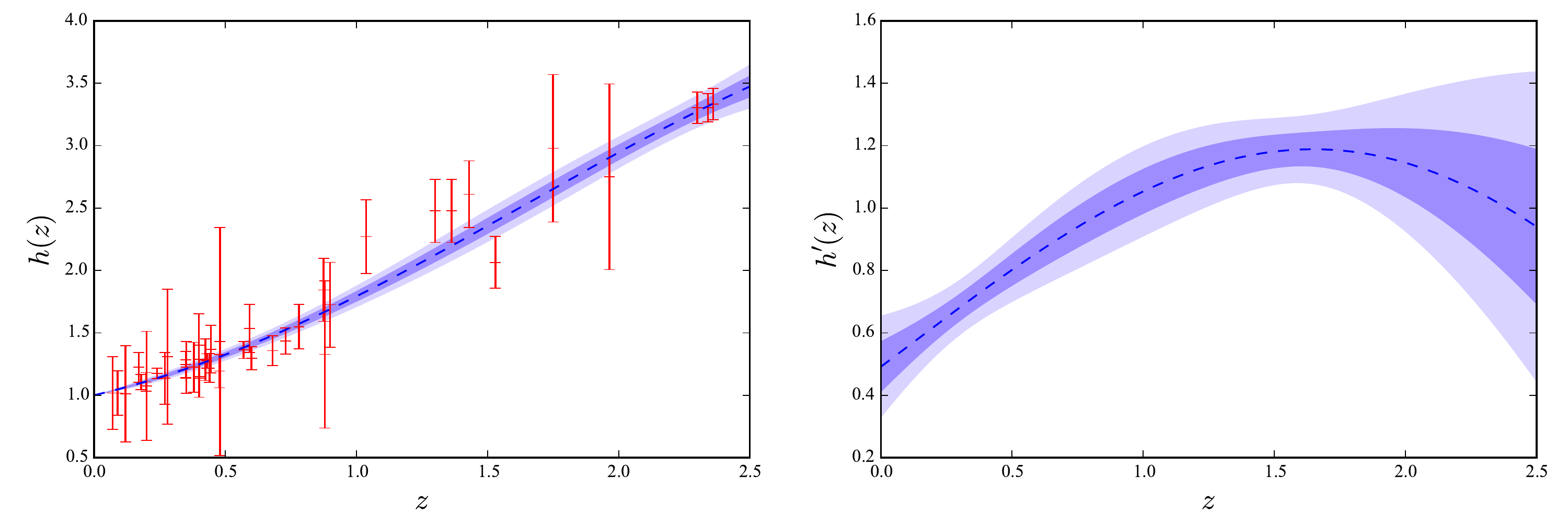}
\caption{The reconstructions of $h(z)$ (left) and $h'(z)$ (right) from $H(z)$ data. The shaded regions represent the $68\%$ and $95\%$ confidence levels. The dashed bule curve is the best fit values of the reconstruction.}\label{hh}
\end{figure}

In this paper, we conveniently use $H(z)$ data sets. There are 36 data points compiled by Meng et al. \citep{meng2015utility}. Among them, 26 data points are deduced from the differential age method, and 10 data points are obtained from the radial BAO method. Besides, just recently, Moresco et al. \citep{Moresco:2016mzx} obtained 5 new data points of $H(z)$ using the differential age method. So, we combine total of 41 data points for the following work. We normalize $H(z)$ using the latest Planck data $H_0=67.8\pm 0.9$ km s$^{-1}$ Mpc$^{-1}$ \citep{planck2015planck}. The uncertainty in $H_0$ is transferred to $h(z)$ as $\sigma_h^2=(\sigma_H^2/H_0^2)+(H^2/H_0^4)\sigma_{H_0}^2$ \citep{seikel2012using}. In addition, we add the theoretical value $h_0=1$ to the data set. The reconstructions of $h(z)$ and $h'(z)$ are shown in Fig. \ref{hh}.

\begin{figure}
\centering
\includegraphics[width=8cm,height=6cm]{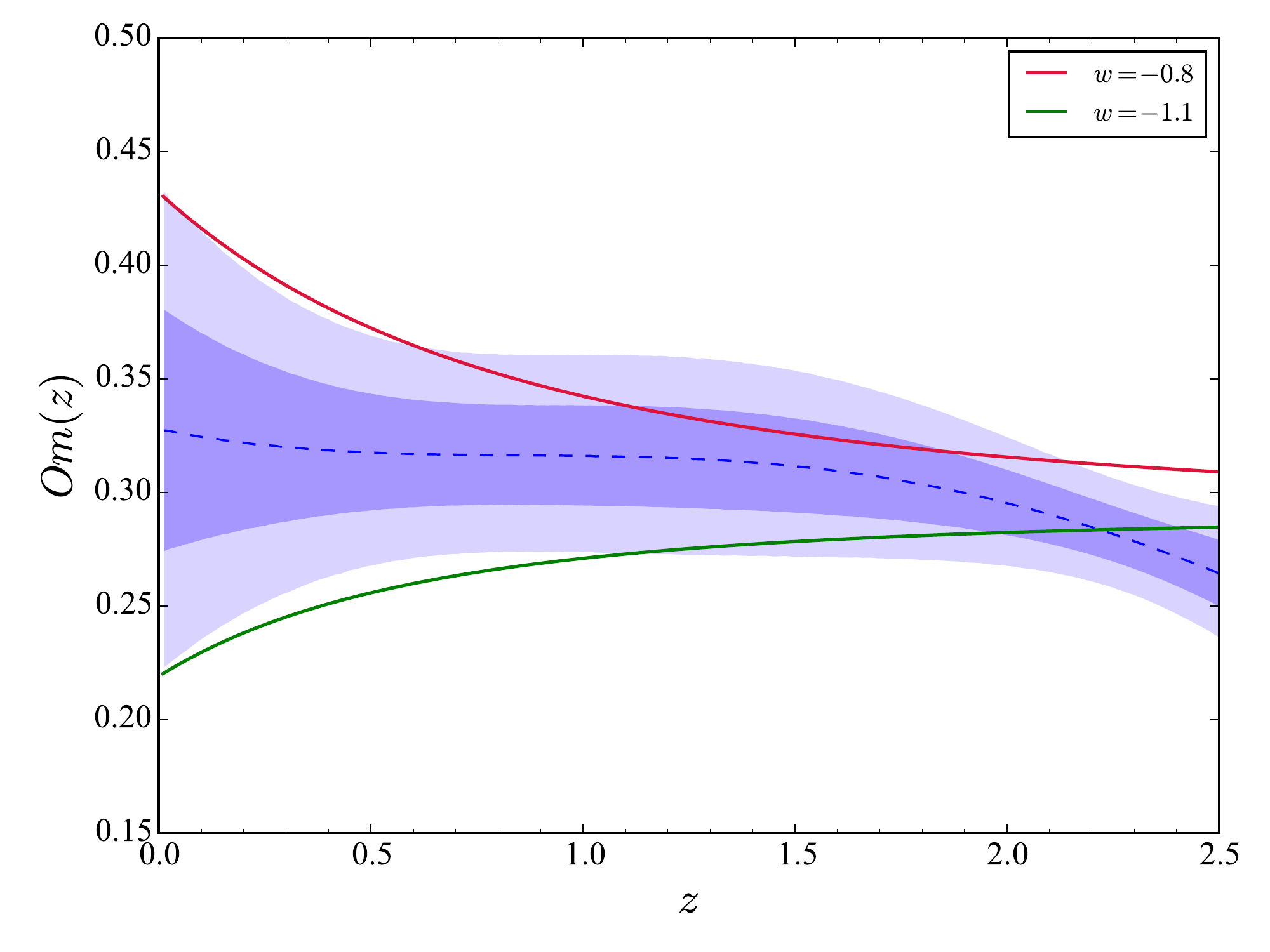}
\caption{The $Om(z)$ reconstructed from $H(z)$ data. The blue dashed line is the best fit reconstruction. Dark and light shaded areas stand for $1 \sigma$ and $2 \sigma$ confidence limits of the reconstructed function, respectively. The red and green line represents the trajectories of $w=-0.8$ and $w=-1.1$, respectively.}\label{oma}
\end{figure}

The reconstructed functions $Om(z)$ is shown in Figs. \ref{oma}. $Om(z)$ has been regarded as a diagnostic to discriminate numerous dark energy models from $\Lambda$CDM. As mentioned above, it works on the principle that different models have different evolutionary trajectories in $z-Om(z)$ plane. If the distance of such trajectories is far enough, it can be said such model could be discriminated in principle. However, the distinguishing criterion is absent in previous works. Theoretically, this criterion should depend on the observational precision. On the other hand, it should be independent of the cosmological model. Actually, the reconstruction of $Om(z)$ here satisfies the two points. In Fig. \ref{oma}, the $Om(z)$ with $1 \sigma$ and $2 \sigma$ confidence are given. If $Om(z)$ of any models go beyond this region, they can be distinguished. The $Om(z)$ is sensitive to EoS of dark energy, namely, a positive slope of $Om(z)$ indicates a phase of phantom ($w<-1$) while a negative slope represents quintessence ($w>-1$). Therefore, we could calculate the differentiated ranges of EoS. As shown in Fig. \ref{oma}, $Om(z)$ can identify the ranges of $w>-0.8$ and $w<-1.1$ at $2 \sigma$ confidence level in low redshift range. However, such differentiated range is still large. It needs the improvement of observational precision.

\begin{figure}
\centering
\includegraphics[width=8cm,height=6cm]{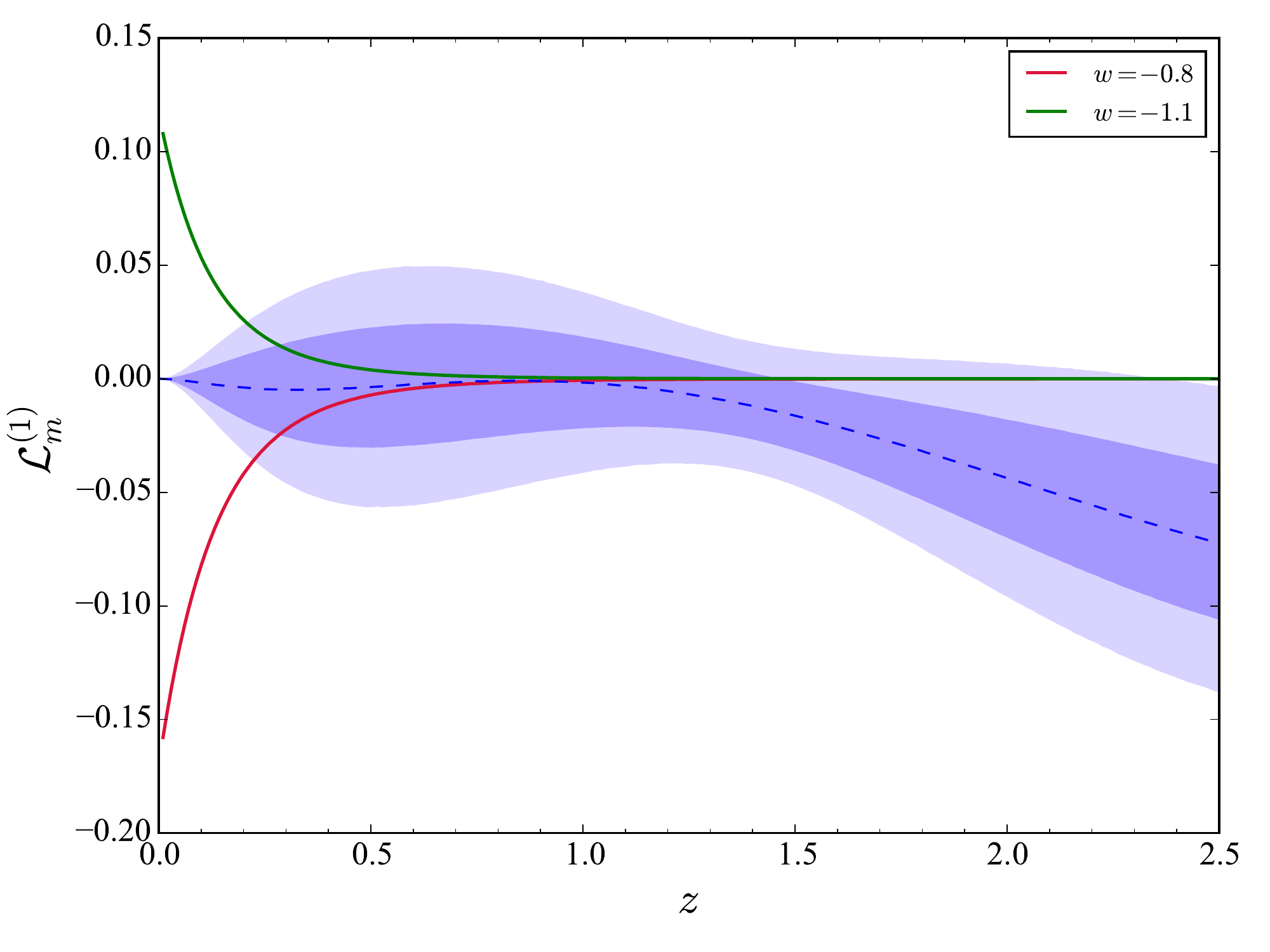}
\caption{The $\mathcal{L}^{(1)}_m(z)$ reconstructed from $H(z)$ data. The blue dashed line is the best fit reconstruction. Dark and light shaded areas stand for $1 \sigma$ and $2 \sigma$ confidence limits of the reconstructed function, respectively. The red and green line represents the trajectories of $w=-0.8$ and $w=-1.1$, respectively.}\label{l1a}
\end{figure}

Fig. \ref{l1a} shows the reconstruction of $\mathcal{L}^{(1)}_m(z)$. As we see, the uncertainty of $\mathcal{L}^{(1)}_m(z)$ is smaller in low redshift range, which means it can present a stronger distinguishing capability. The $\mathcal{L}^{(1)}_m(z)$ curves of $w=-0.8$ and $w=-1.1$ are far beyond the $2 \sigma$ of reconstruction of $\mathcal{L}^{(1)}_m(z)$ in low redshift range. The $\mathcal{L}^{(1)}_m(z)$ is expected to discriminate models and estimate the authenticity of various models.

\section{Dark energy models} \label{DE}
As mentioned before, more and more dark energy models have been proposed. We could use the above reconstructions to discriminate some models and test their authenticity. Here, we will focus on some popular models, such as $\Lambda$CDM, GCG, CPL and JBP models. These models and their discriminations by Statefinder hierarchy have be discussed in detail in Refs. \citep{Li:2014mua,qi2015several}. As follows, we adopt a spatial flat FRW universe and just consider the contribution of dark energy and matter.

(1) The $\Lambda$CDM model is the most robust model. For a flat space, it only has one free parameter $\Omega_{m0}$. Its normalized Hubble parameter is
\begin{eqnarray}
E(z)=\frac{H(z)}{H_0}=\left[\Omega_{m0}(1+z)^3+(1-\Omega_{m0})\right]^{1/2}.
\end{eqnarray}
According to nine-year Wilkinson Microwave Anisotropy Probe (WMAP) observations \citep{hinshaw2013nine}, we take $\Omega_{m0}=0.2855^{+0.0096}_{-0.0097}$.

(2) The generalized Chaplygin gas (GCG) has the generalized EoS $p=-A/\rho^{\alpha}$ and $0<\alpha \leq 1$, where $A$ is a positive constant. The GCG model is a unified dark matter and dark energy model. Introducing $A_s=A/\rho_0^{1+\alpha}$ with the present value of the energy density of GCG $\rho_0$, the EoS and the normalized Hubble parameter of GCG could be expressed as
\begin{eqnarray}
w&=&-\frac{A_s}{A_s+(1-A_s)(1+z)^{3(1+\alpha)}}, \\
E(z)&=&\frac{H(z)}{H_0}=\left[\Omega_{b0}(1+z)^3+(1-\Omega_{b0})\left( A_s+(1-A_s)(1+z)^{3(1+\alpha)}\right) ^{\frac{1}{1+\alpha}}\right]^{1/2},
\end{eqnarray}
respectively. The constraints of $\Omega_{m0}$, $A_s$ and $\alpha$ with $1 \sigma$ and $2 \sigma$ could be found in Ref. \citep{xu2010cosmological}. The values of these parameters could be took as $\Omega_{b0}=0.0233^{+0.0023}_{-0.0016}$, $A_s=0.760^{+0.029}_{-0.039}$ and $\alpha=0.033^{+0.066}_{-0.071}$.

(3) On the other hand, there are a lot of parameterizations for the EoS of dark energy, which have been widely employed to analyse the behavior of dark energy. The most popular parameterization is Chevallier-Polarski-Linder (CPL) parametrization \citep{chevallier2001accelerating,linder2003exploring}:
\begin{equation}
w_{de}=w_0+w_a(1-a)=w_0+w_a\frac{z}{1+z}, \label{cpl}
\end{equation}
where $w_0$ and $w_a$ are constants. Its normalized Hubble parameter for a flat universe is
\begin{eqnarray}
E^2(z)&=&\frac{H^2(z)}{H^{2}_0} =\Omega_{m0}(1+z)^3+(1-\Omega_{m0})(1+z)^{3(1+w_0+w_a)}\exp\left(\frac{-3w_az}{1+z} \right). \label{cplE}
\end{eqnarray}
According to nine-year Wilkinson Microwave Anisotropy Probe (WMAP) observations \citep{hinshaw2013nine}, we take $\Omega_{m0}=0.2855^{+0.0096}_{-0.0097}$, $w_0=-1.17^{+0.13}_{-0.12}$ and $w_a=0.35^{+0.50}_{-0.49}$.

(4) Another popular parameterization, Jassal-Bagla-Padmanabhan (JBP), is also studied here. Its EoS and normalized Hubble parameter take the form
\begin{eqnarray}
w_{de}&=&w_0+w_a\frac{z}{(1+z)^2}, \label{JBP} \\
E^2(z)&=&\Omega_{m0}(1+z)^3+(1-\Omega_{m0})(1+z)^{3(1+w_0)}\exp\left(\frac{3w_az^2}{2(1+z)^2} \right), \label{jpbE}
\end{eqnarray}
where $w_0$ and $w_a$ are constants. According Ref. \citep{shi2011effects}, the values of such parameters could be took as $\Omega_{m0}=0.28^{+0.01}_{-0.01}$, $w_0=-1.03^{+0.10}_{-0.10}$ and $w_a=0.95^{+0.92}_{-0.84}$.

\section{Results and discussions}\label{RD}
\begin{figure}
\centering
\includegraphics[width=7cm,height=5cm]{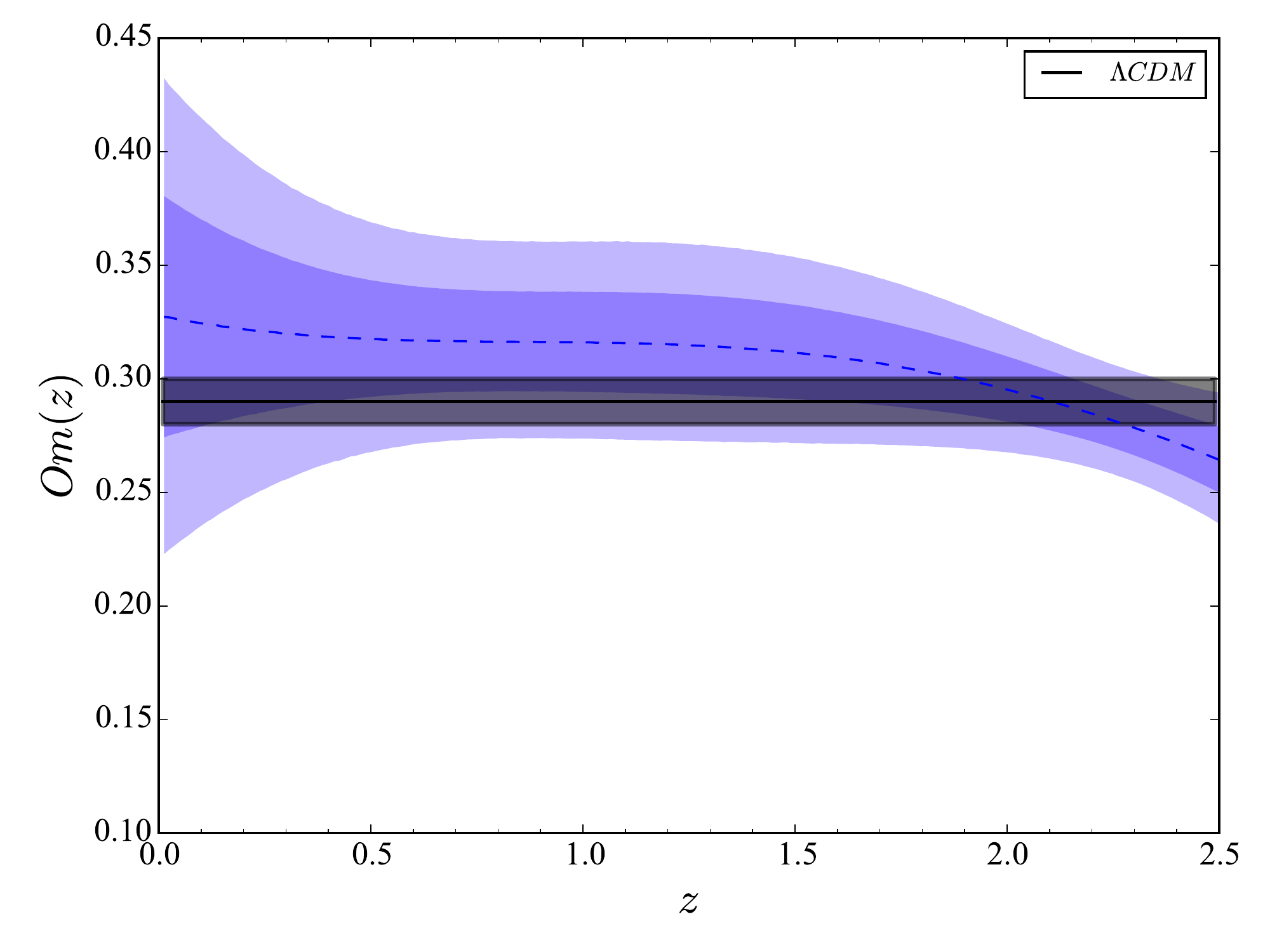}
\includegraphics[width=7cm,height=5cm]{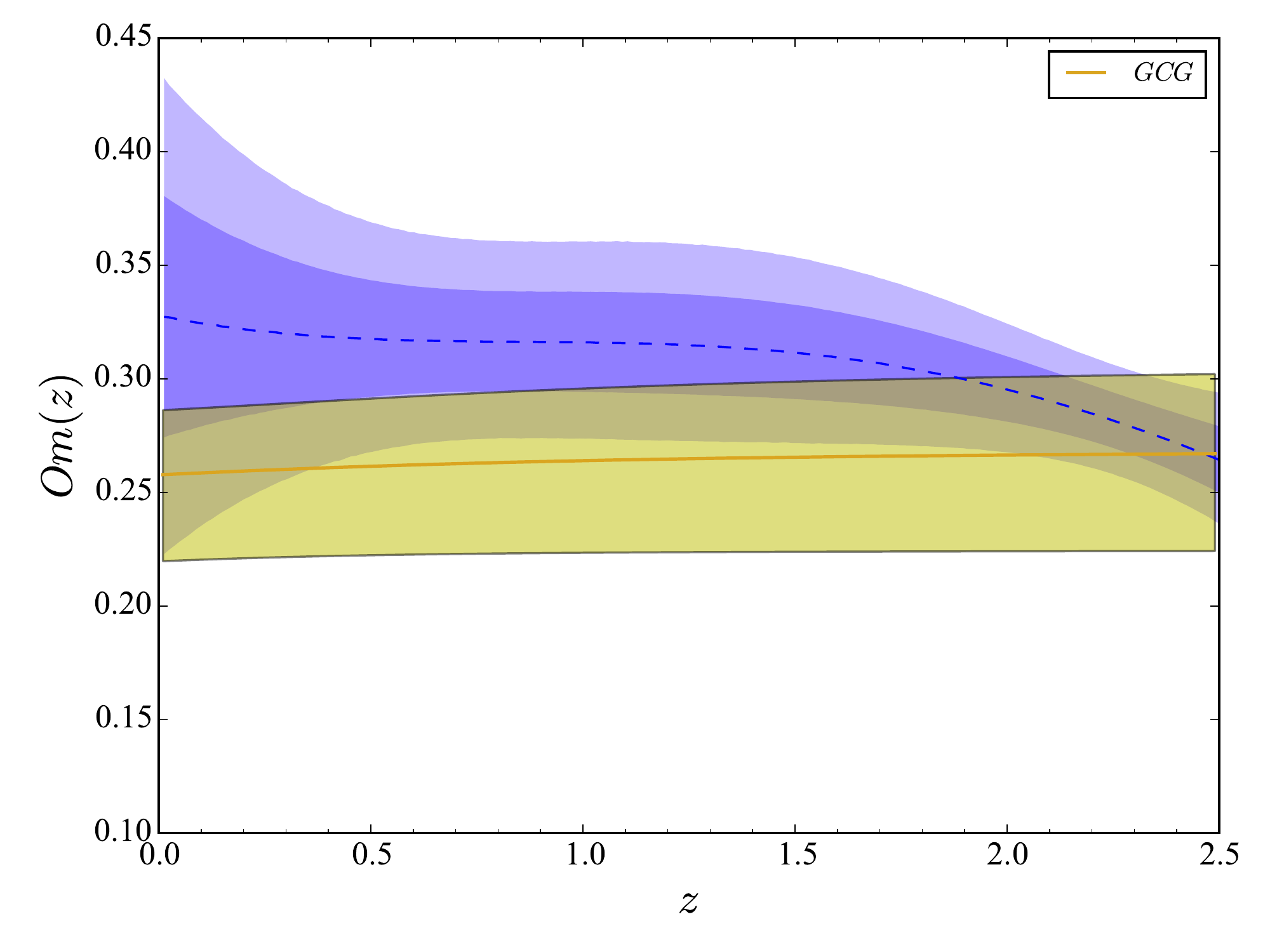}
\includegraphics[width=7cm,height=5cm]{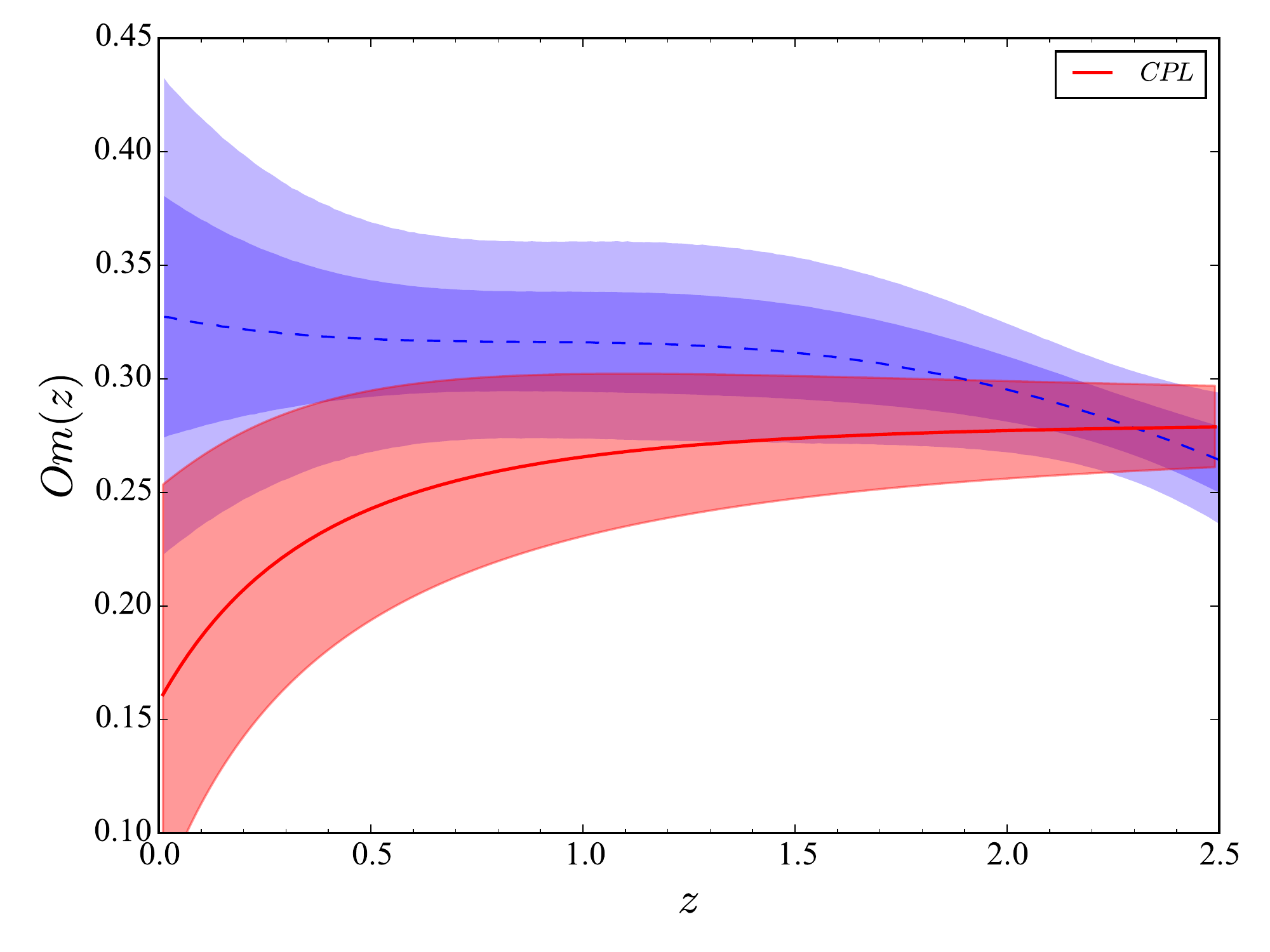}
\includegraphics[width=7cm,height=5cm]{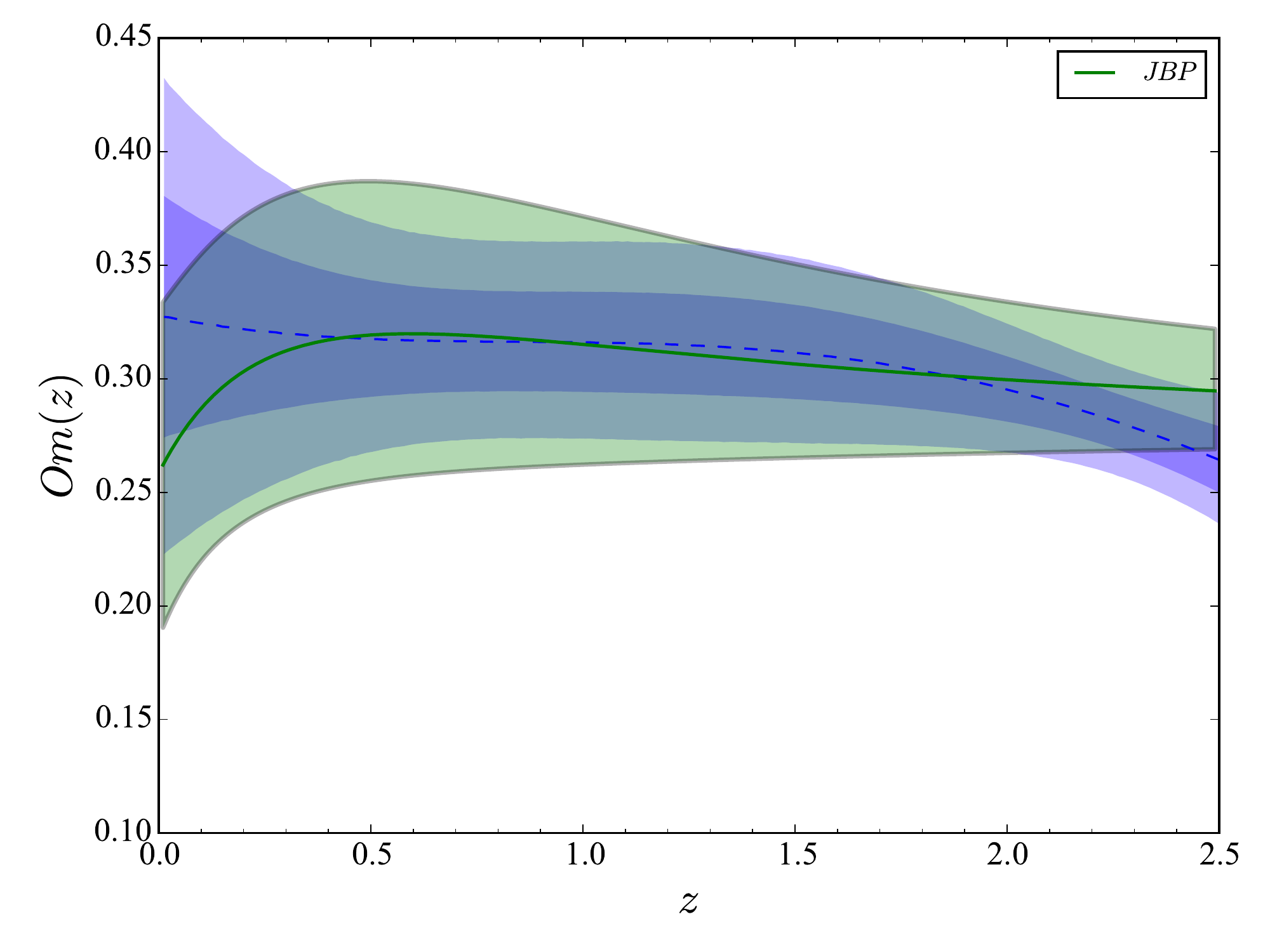}
\caption{The $Om(z)$ of reconstruction and theoretical values with $1 \sigma$ confidence range from $\Lambda$CDM, GCG, CPL and JBP.}\label{omc}
\end{figure}

\begin{figure}
\centering
\includegraphics[width=7cm,height=5cm]{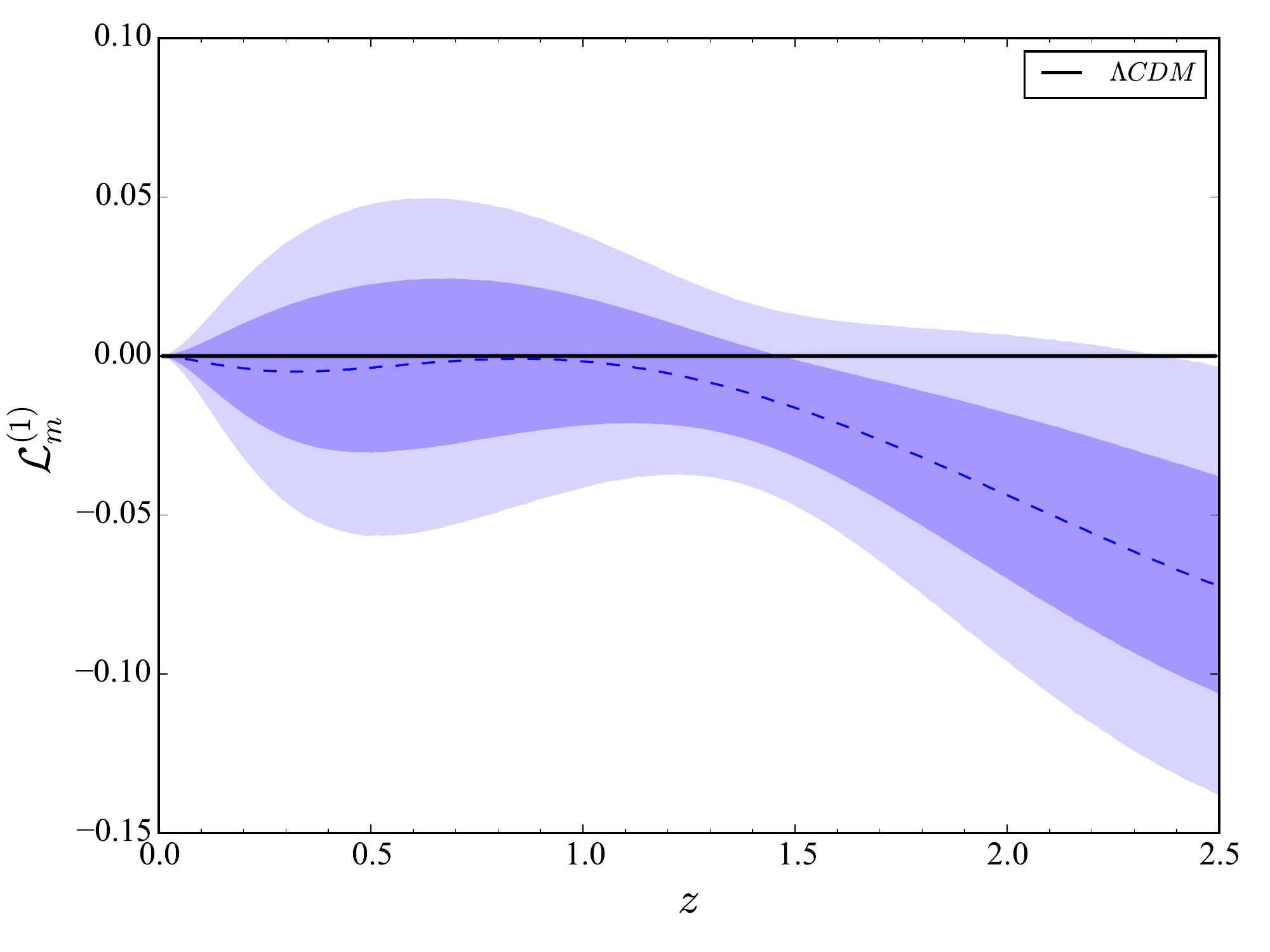}
\includegraphics[width=7cm,height=5cm]{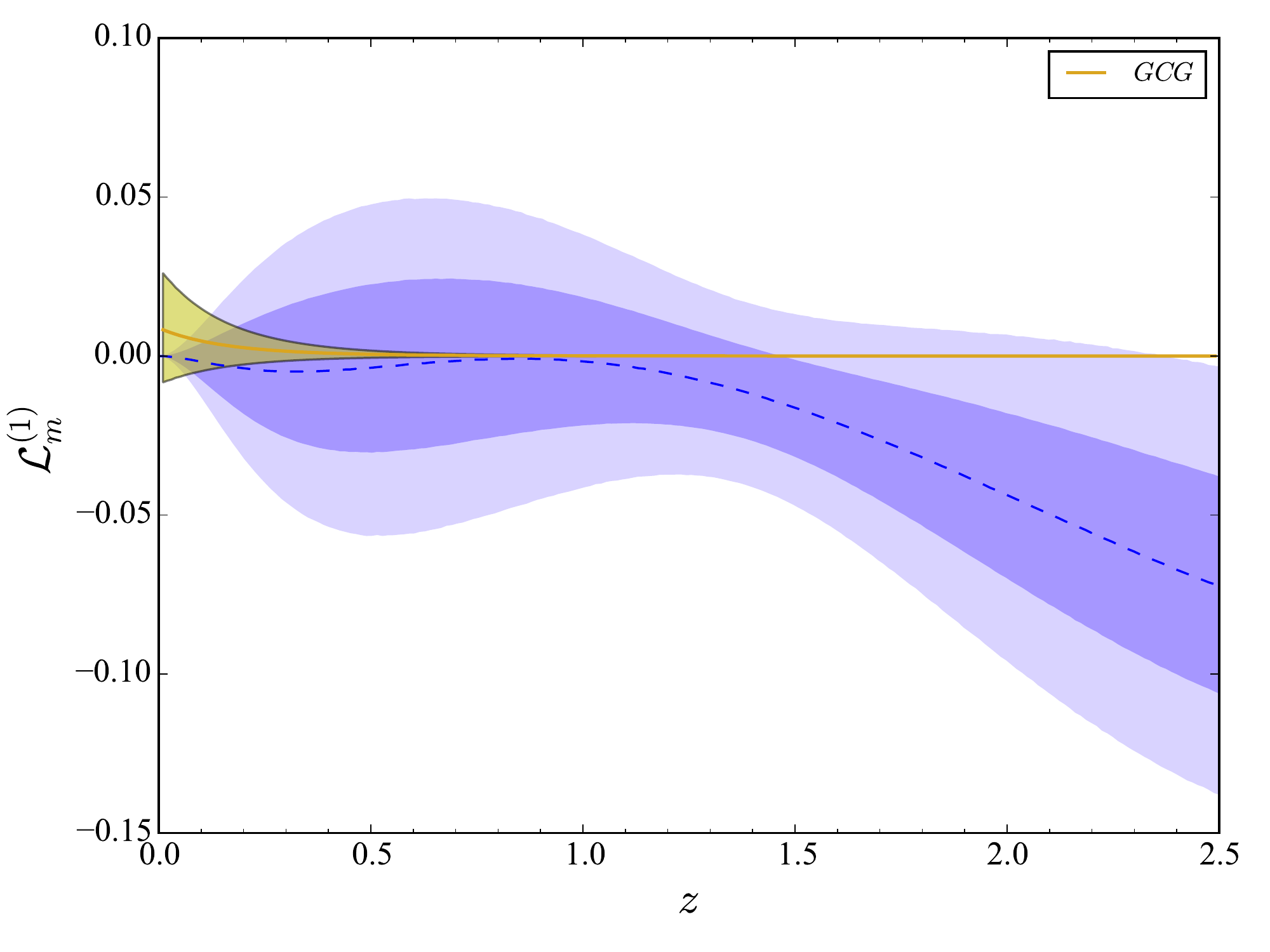}
\includegraphics[width=7cm,height=5cm]{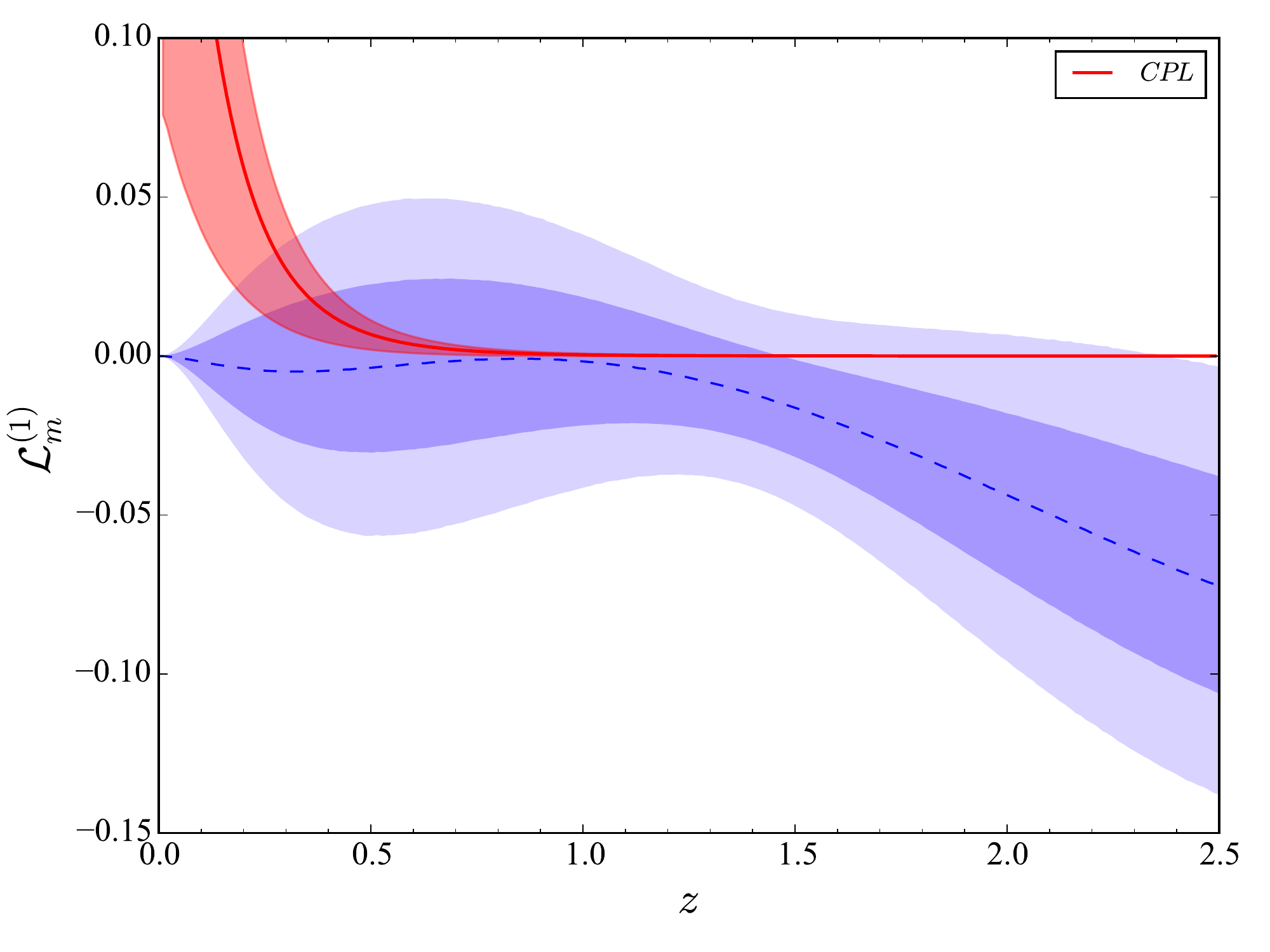}
\includegraphics[width=7cm,height=5cm]{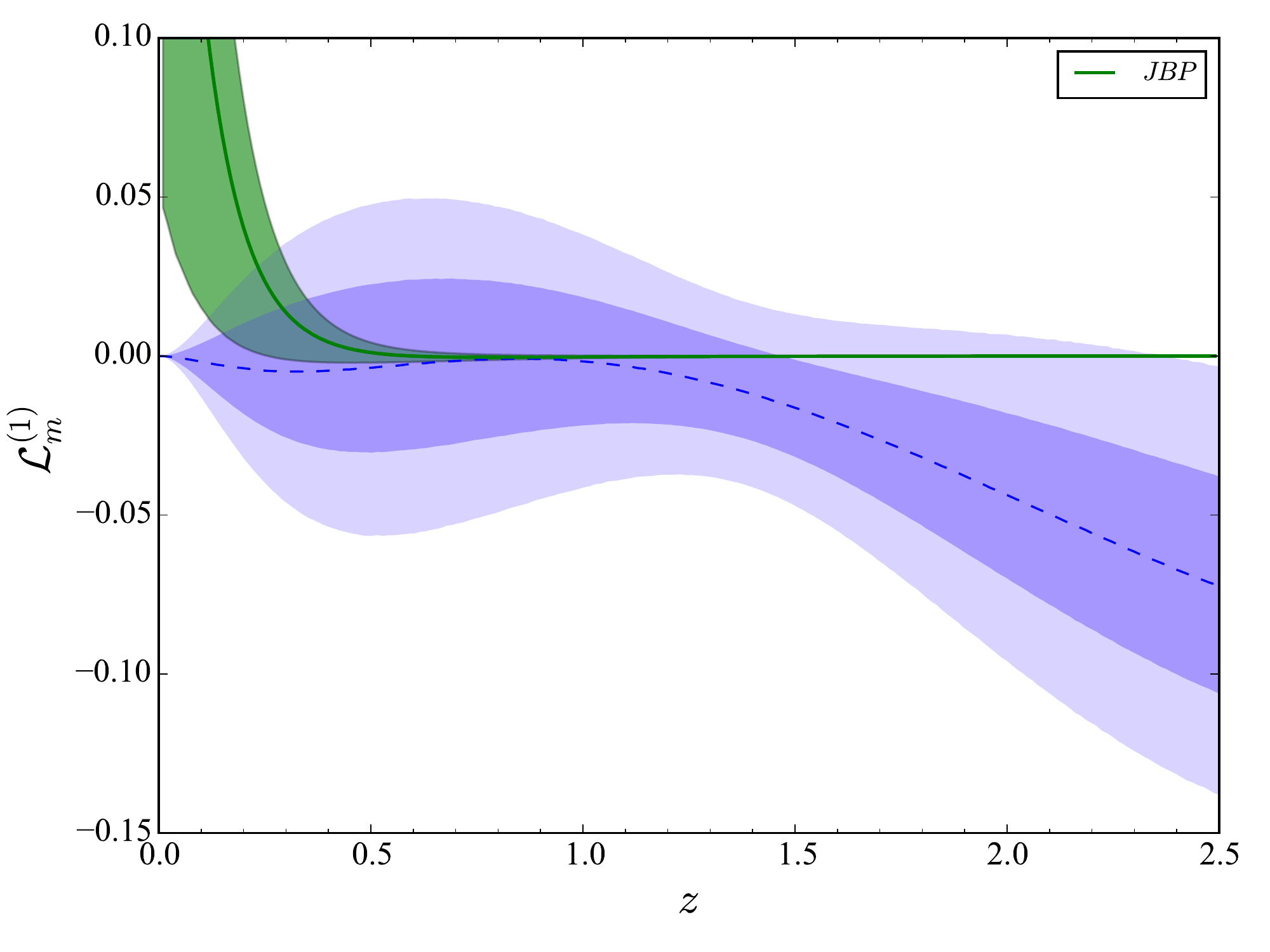}
\caption{The $\mathcal{L}^{(1)}_m$ of reconstruction and theoretical values with $1 \sigma$ confidence range from $\Lambda$CDM, GCG, CPL and JBP.}\label{wucha}
\end{figure}

The values of the parameters of these dark energy models have been given, and then their evolutions can also be obtained. Considering $1 \sigma$ error of parameters of the models, we plot their $Om(z)$ with error ranges using constructing method in Refs. \citep{qi2016transient,lazkoz2012first}, and then compare them with Fig. \ref{oma} reconstructed by $H(z)$ data. The results are shown in Fig. \ref{omc}. The $1 \sigma$ confidence range of $\Lambda CDM$ and JBP models almost overlap completely with the $1 \sigma$ confidence range of $Om(z)$ reconstructed by $H(z)$ data, which means they can not be distinguished or ruled out by $H(z)$ data. The $1 \sigma$ confidence range of GCG model overlaps slightly with the $1 \sigma$ confidence range of reconstructed $Om(z)$ in low redshift range.
The $1 \sigma$ confidence range of CPL model has some overlaps with the $2 \sigma$ confidence range of reconstructed $Om(z)$, however, it has deviated from the $1 \sigma$ confidence range of reconstructed $Om(z)$ in low redshift. This result indicates that the CPL model is not reliable enough for $H(z)$ data at $1 \sigma$ confidence level.

As mentioned above, $\mathcal{L}^{(1)}_m$ has more stronger distinguishing capability, so we calculate $\mathcal{L}^{(1)}_m$ of the models and compare them with the reconstruction (see Fig. \ref{wucha}). $\Lambda CDM$ model is still the best model in accordance with observation. Its $1 \sigma$ confidence range is so small that the error area could not be seen. The $1 \sigma$ confidence range of GCG model almost completely contains or is contained with the $1 \sigma$ confidence range of $\mathcal{L}^{(1)}_m$ reconstructed by $H(z)$ data. In other words, the reconstructed $\mathcal{L}^{(1)}_m$ could not distinguish or rule out GCG model. While, the $1 \sigma$ confidence range of CPL and JBP models has visible deviations from the $1 \sigma$ confidence range of reconstructed $\mathcal{L}^{(1)}_m$ in low redshift. That is to say, the $H(z)$ data does not favor the CPL and JBP models at $1 \sigma$ confidence level.

More interestingly, in high redshift range, the trajectories of all these models not only have deviated from the reconstructed $\mathcal{L}^{(1)}_m$ with $1 \sigma$ confidence level, but also have a tendency to depart from $2 \sigma$ confidence level. We also note that the reconstructed $\mathcal{L}^{(1)}_m$ gradually deviates from zero at $z>1.5$. Note, these deviations have been considered $1 \sigma$ error of parameters of models, which means the theoretical values disagree with observed values at $68.3\%$ confidence level in high redshift range. We think there are two possible explanations for this unusual deviation. One possibility is that high redshift data are thin, so the results are not accurate enough. Another possibility is that this tendency is authentic, and then the models considered are unreliable. Actually, using an improved version of the $Om$ diagnostic, Sahni et al. \citep{sahni2014model} demonstrated that one high redshift data, $H(z=2.34)$ from BAO data, is disagreeable with standard $\Lambda$CDM. And using a data set including 29 $H(z)$, Ding et al. \citep{ding2015there} further confirmed this discrepancy not only for $\Lambda$CDM model but also other dark energy models based on general relativity. So, the second explanation sounds reasonable. Our results also demonstrate these models are disagreeable with high redshift $H(z)$ data, which supports the conclusions of Sahni et al. \citep{sahni2014model} and Ding et al. \citep{ding2015there} that $\Lambda$CDM model may not be the best description of our universe. Anyway, it needs more high redshift data to verify. If the future data supports this unusual deviation, the present theories will face a great challenge.

\section{Conclusion}\label{Con}

As more and more dark energy models were proposed, some diagnostics consequently were also proposed to distinguish these increasing models. However, most of the diagnostics cannot directly compare with the observational data. Therefore, they cannot estimate which model is more realistic. In this paper, we focus on the $Om(z)$ and its fist derivative $\mathcal{L}^{(1)}_m$. $Om(z)$ has been regarded as a diagnostic to discriminate numerous dark energy models from $\Lambda$CDM. But there were few works to discuss the distinguishing criterion. We reconstruct the $Om(z)$ and  $\mathcal{L}^{(1)}_m$ from the latest observational $H(z)$ data, which could be used as the distinguishing criterion. Our results indicate $Om(z)$ could identify the ranges of $w>-0.8$ and $w<-1.1$ at $2\sigma$ confidence level in low redshift range. In addition, we find that $\mathcal{L}^{(1)}_m$ has a stronger distinguishing capability in low redshift range. These two quantities are expected to discriminate models and judge the authenticity of various models. We choose some popular models to study, such as $\Lambda$CDM, GCG, CPL and JBP.

Finally, we plot the trajectories of $Om(z)$ and $\mathcal{L}^{(1)}_m$ with $1 \sigma$ confidence level of these models, and compare them to the reconstruction from $H(z)$ data set. The results show that $Om(z)$ cannot distinguish $\Lambda$CDM, GCG and JBP at $1 \sigma$ confidence level. The $1 \sigma$ confidence range of CPL model has some overlaps with the $2 \sigma$ confidence range of reconstructed $Om(z)$, however, it has deviated from the $1 \sigma$ confidence range of reconstructed $Om(z)$ in low redshift. This result indicates that the CPL model is not reliable enough for $H(z)$ data at $1 \sigma$ confidence level.

In the $z-\mathcal{L}^{(1)}_m$ plane, the $1 \sigma$ confidence range of GCG and $\Lambda$CDM models almost completely contain or is contained with the $1 \sigma$ confidence range of $\mathcal{L}^{(1)}_m$ reconstructed by $H(z)$ data. While, the $1 \sigma$ confidence range of CPL and JBP models has visible deviations from the $1 \sigma$ confidence range of reconstructed $\mathcal{L}^{(1)}_m$ in low redshift. That is to say, the $H(z)$ data does not favor the CPL and JBP models at $1 \sigma$ confidence level.

Strangely, in high redshift range, the reconstructed $\mathcal{L}^{(1)}_m$ has a tendency of deviation from theoretical value, which demonstrates these models are all disagreeable with high redshift $H(z)$ data. This result supports the conclusions of Sahni et al. \citep{sahni2014model} and Ding et al. \citep{ding2015there} that $\Lambda$CDM model may not be the best description of our universe. Anyway, it needs more high redshift data to verify. If the future data supports this unusual deviation, the present theories will face a great challenge.

\acknowledgments
J.-Z. Qi would like to express his gratitude towards PhD. Tao Yang for his generous help. This work is supported by the National Natural Science Foundation of China (Grant Nos. 11235003, 11175019, and 11178007). M.-J. Zhang is funded by China Postdoctoral Science Foundation under Grant No. 2015M581173.

\bibliographystyle{JHEP}
\bibliography{Notes}

\providecommand{\href}[2]{#2}\begingroup\raggedright\begin{thebibliography}{10}

\bibitem{sahni2014model}
V.~Sahni, A.~Shafieloo, and A.~A. Starobinsky, {\it Model-independent evidence
  for dark energy evolution from baryon acoustic oscillations},  {\em The
  Astrophysical Journal Letters} {\bf 793} (2014), no.~2 L40.

\bibitem{ding2015there}
X.~Ding, M.~Biesiada, S.~Cao, Z.~Li, and Z.-H. Zhu, {\it Is there evidence for
  dark energy evolution?},  {\em The Astrophysical Journal Letters} {\bf 803}
  (2015), no.~2 L22.

\bibitem{riess1998supernova}
A.~Riess et~al., {\it Supernova search team collaboration},  {\em Astron. J}
  {\bf 116} (1998) 1009.

\bibitem{tegmark2004cosmological}
M.~Tegmark, M.~Strauss, M.~Blanton, et~al., {\it Cosmological parameters from
  sdss and wmap},  {\em Physical Review D} {\bf 69} (2004), no.~10 103501.

\bibitem{spergel2003wmap}
D.~Spergel et~al., {\it Wmap collaboration},  {\em Astrophys. J. Suppl} {\bf
  148} (2003), no.~175 170.

\bibitem{weinberg1989cosmological}
S.~Weinberg, {\it The cosmological constant probiem},  {\em Rev. Mod. Phys}
  {\bf 61} (1989), no.~1.

\bibitem{weinberg2000cosmological}
S.~Weinberg, {\it The cosmological constant problems (talk given at dark matter
  2000, february, 2000)},  {\em arXiv preprint astro-ph/0005265} (2000).

\bibitem{1999PhRvL..82..896Z}
I.~{Zlatev}, L.~{Wang}, and P.~J. {Steinhardt}, {\it {Quintessence, Cosmic
  Coincidence, and the Cosmological Constant}},  {\em Physical Review Letters}
  {\bf 82} (Feb., 1999) 896--899,
  [\href{http://arxiv.org/abs/astro-ph/9807002}{{\tt astro-ph/9807002}}].

\bibitem{Caldwell:2005tm}
R.~Caldwell and E.~V. Linder, {\it {The Limits of quintessence}},  {\em
  Phys.Rev.Lett.} {\bf 95} (2005) 141301,
  [\href{http://arxiv.org/abs/astro-ph/0505494}{{\tt astro-ph/0505494}}].

\bibitem{Zlatev:1998tr}
I.~Zlatev, L.-M. Wang, and P.~J. Steinhardt, {\it {Quintessence, cosmic
  coincidence, and the cosmological constant}},  {\em Phys.Rev.Lett.} {\bf 82}
  (1999) 896--899, [\href{http://arxiv.org/abs/astro-ph/9807002}{{\tt
  astro-ph/9807002}}].

\bibitem{Tsujikawa:2013fta}
S.~Tsujikawa, {\it {Quintessence: A Review}},  {\em Class.Quant.Grav.} {\bf 30}
  (2013) 214003, [\href{http://arxiv.org/abs/1304.1961}{{\tt
  arXiv:1304.1961}}].

\bibitem{chiba2000kinetically}
T.~Chiba, T.~Okabe, and M.~Yamaguchi, {\it Kinetically driven quintessence},
  {\em Physical Review D} {\bf 62} (2000), no.~2 023511.

\bibitem{armendariz2000dynamical}
C.~Armendariz-Picon, V.~Mukhanov, and P.~J. Steinhardt, {\it Dynamical solution
  to the problem of a small cosmological constant and late-time cosmic
  acceleration},  {\em Physical Review Letters} {\bf 85} (2000), no.~21 4438.

\bibitem{kahya2007quantum}
E.~O. Kahya and V.~K. Onemli, {\it Quantum stability of a w<- 1 phase of cosmic
  acceleration},  {\em Physical Review D} {\bf 76} (2007), no.~4 043512.

\bibitem{onemli2004quantum}
V.~Onemli and R.~Woodard, {\it Quantum effects can render w<- 1 on cosmological
  scales},  {\em Physical Review D} {\bf 70} (2004), no.~10 107301.

\bibitem{singh2003cosmological}
P.~Singh, M.~Sami, and N.~Dadhich, {\it Cosmological dynamics of a phantom
  field},  {\em Physical Review D} {\bf 68} (2003), no.~2 023522.

\bibitem{bento2002generalized}
M.~Bento, O.~Bertolami, and A.~Sen, {\it Generalized chaplygin gas, accelerated
  expansion, and dark-energy-matter unification},  {\em Physical Review D} {\bf
  66} (2002), no.~4 043507.

\bibitem{kamenshchik2001alternative}
A.~Kamenshchik, U.~Moschella, and V.~Pasquier, {\it An alternative to
  quintessence},  {\em Physics Letters B} {\bf 511} (2001), no.~2 265--268.

\bibitem{riess2004type}
A.~G. Riess, L.-G. Strolger, J.~Tonry, S.~Casertano, H.~C. Ferguson,
  B.~Mobasher, P.~Challis, A.~V. Filippenko, S.~Jha, W.~Li, et~al., {\it Type
  ia supernova discoveries at z> 1 from the hubble space telescope: Evidence
  for past deceleration and constraints on dark energy evolution},  {\em The
  Astrophysical Journal} {\bf 607} (2004), no.~2 665.

\bibitem{barboza2009generalized}
E.~Barboza~Jr, J.~Alcaniz, Z.-H. Zhu, and R.~Silva, {\it Generalized equation
  of state for dark energy},  {\em Physical Review D} {\bf 80} (2009), no.~4
  043521.

\bibitem{zhang2015exploring}
Q.~Zhang, G.~Yang, Q.~Zou, X.~Meng, and K.~Shen, {\it Exploring the low
  redshift universe: two parametric models for effective pressure},  {\em The
  European Physical Journal C} {\bf 75} (2015), no.~7 300.

\bibitem{maor2001limitations}
I.~Maor, R.~Brustein, and P.~J. Steinhardt, {\it Limitations in using
  luminosity distance to determine the equation of state of the universe},
  {\em Physical Review Letters} {\bf 86} (2001), no.~1 6.

\bibitem{chevallier2001accelerating}
M.~Chevallier and D.~Polarski, {\it Accelerating universes with scaling dark
  matter},  {\em International Journal of Modern Physics D} {\bf 10} (2001),
  no.~02 213--223.

\bibitem{linder2003exploring}
E.~V. Linder, {\it Exploring the expansion history of the universe},  {\em
  Physical Review Letters} {\bf 90} (2003), no.~9 091301.

\bibitem{Jassal:2004ej}
H.~K. Jassal, J.~S. Bagla, and T.~Padmanabhan, {\it {WMAP constraints on low
  redshift evolution of dark energy}},  {\em Mon. Not. Roy. Astron. Soc.} {\bf
  356} (2005) L11--L16, [\href{http://arxiv.org/abs/astro-ph/0404378}{{\tt
  astro-ph/0404378}}].

\bibitem{Wei:2013jya}
H.~Wei, X.-P. Yan, and Y.-N. Zhou, {\it {Cosmological Applications of Pad¨¦
  Approximant}},  {\em JCAP} {\bf 1401} (2014) 045,
  [\href{http://arxiv.org/abs/1312.1117}{{\tt arXiv:1312.1117}}].

\bibitem{Sahni:2002fz}
V.~Sahni, T.~D. Saini, A.~A. Starobinsky, and U.~Alam, {\it {Statefinder: A New
  geometrical diagnostic of dark energy}},  {\em JETP Lett.} {\bf 77} (2003)
  201--206, [\href{http://arxiv.org/abs/astro-ph/0201498}{{\tt
  astro-ph/0201498}}]. [Pisma Zh. Eksp. Teor. Fiz.77,249(2003)].

\bibitem{sahni2008two}
V.~Sahni, A.~Shafieloo, and A.~A. Starobinsky, {\it Two new diagnostics of dark
  energy},  {\em Physical Review D} {\bf 78} (2008), no.~10 103502.

\bibitem{Arabsalmani:2011fz}
M.~Arabsalmani and V.~Sahni, {\it {The Statefinder hierarchy: An extended null
  diagnostic for concordance cosmology}},  {\em Phys. Rev.} {\bf D83} (2011)
  043501, [\href{http://arxiv.org/abs/1101.3436}{{\tt arXiv:1101.3436}}].

\bibitem{shojai2009statefinder}
A.~Shojai and F.~Shojai, {\it Statefinder diagnosis of nearly flat and thawing
  non-minimal quintessence},  {\em EPL (Europhysics Letters)} {\bf 88} (2009),
  no.~3 30002.

\bibitem{zhang2005statefinder}
X.~Zhang, {\it Statefinder diagnostic for coupled quintessence},  {\em Physics
  Letters B} {\bf 611} (2005), no.~1 1--7.

\bibitem{bao2007statefinder}
C.~Bao-Rong, L.~Hong-Ya, X.~Li-Xin, and Z.~Cheng-Wu, {\it Statefinder
  diagnostic for phantom model with v (phi)= v0exp (- $\lambda$phi2)},  {\em
  Chinese Physics Letters} {\bf 24} (2007), no.~7 2153.

\bibitem{chang2007statefinder}
B.~Chang, H.~Liu, L.~Xu, C.~Zhang, and Y.~Ping, {\it Statefinder parameters for
  interacting phantom energy with dark matter},  {\em Journal of Cosmology and
  Astroparticle Physics} {\bf 2007} (2007), no.~01 016.

\bibitem{feng2008statefinder}
C.-J. Feng, {\it Statefinder diagnosis for ricci dark energy},  {\em Physics
  Letters B} {\bf 670} (2008), no.~3 231--234.

\bibitem{Li:2014mua}
J.~Li, R.~Yang, and B.~Chen, {\it {Discriminating dark energy models by using
  the statefinder hierarchy and the growth rate of matter perturbations}},
  {\em JCAP} {\bf 1412} (2014), no.~12 043,
  [\href{http://arxiv.org/abs/1406.7514}{{\tt arXiv:1406.7514}}].

\bibitem{qi2015several}
J.-Z. Qi and W.-B. Liu, {\it Several parametrization dark energy models
  comparison with statefinder hierarchy},  {\em arXiv preprint
  arXiv:1510.02633} (2015).

\bibitem{seikel2012using}
M.~Seikel, S.~Yahya, R.~Maartens, and C.~Clarkson, {\it Using h (z) data as a
  probe of the concordance model},  {\em Physical Review D} {\bf 86} (2012),
  no.~8 083001.

\bibitem{shafieloo2010model}
A.~Shafieloo and C.~Clarkson, {\it Model independent tests of the standard
  cosmological model},  {\em Physical Review D} {\bf 81} (2010), no.~8 083537.

\bibitem{seikel2012reconstruction}
M.~Seikel, C.~Clarkson, and M.~Smith, {\it Reconstruction of dark energy and
  expansion dynamics using gaussian processes},  {\em Journal of Cosmology and
  Astroparticle Physics} {\bf 2012} (2012), no.~06 036.

\bibitem{gapp}
\url{http://www.acgc.uct.ac.za/~seikel/GAPP/index.html}.

\bibitem{yahya2014null}
S.~Yahya, M.~Seikel, C.~Clarkson, R.~Maartens, and M.~Smith, {\it Null tests of
  the cosmological constant using supernovae},  {\em Physical Review D} {\bf
  89} (2014), no.~2 023503.

\bibitem{cai2015reconstructing}
R.-G. Cai, Z.-K. Guo, and T.~Yang, {\it Reconstructing interaction between dark
  energy and dark matter using gaussian processes},  {\em arXiv preprint
  arXiv:1505.04443} (2015).

\bibitem{seikel2013optimising}
M.~Seikel and C.~Clarkson, {\it Optimising gaussian processes for
  reconstructing dark energy dynamics from supernovae},  {\em arXiv preprint
  arXiv:1311.6678} (2013).

\bibitem{cai2015null}
R.-G. Cai, Z.-K. Guo, and T.~Yang, {\it Null test of the cosmic curvature using
  $ h (z) $ and supernovae data},  {\em arXiv preprint arXiv:1509.06283}
  (2015).

\bibitem{meng2015utility}
X.-L. Meng, X.~Wang, S.-Y. Li, and T.-J. Zhang, {\it Utility of observational
  hubble parameter data on dark energy evolution},  {\em arXiv preprint
  arXiv:1507.02517} (2015).

\bibitem{Moresco:2016mzx}
M.~Moresco, L.~Pozzetti, A.~Cimatti, R.~Jimenez, C.~Maraston, L.~Verde,
  D.~Thomas, A.~Citro, R.~Tojeiro, and D.~Wilkinson, {\it {A 6\% measurement of
  the Hubble parameter at $z\sim0.45$: direct evidence of the epoch of cosmic
  re-acceleration}},  \href{http://arxiv.org/abs/1601.01701}{{\tt
  arXiv:1601.01701}}.

\bibitem{planck2015planck}
P.~Collaboration et~al., {\it Planck 2015 results. xiii. cosmological
  parameters},  {\em arXiv preprint arXiv:1502.01589} (2015).

\bibitem{hinshaw2013nine}
G.~Hinshaw, D.~Larson, E.~Komatsu, D.~Spergel, C.~Bennett, J.~Dunkley,
  M.~Nolta, M.~Halpern, R.~Hill, N.~Odegard, et~al., {\it Nine-year wilkinson
  microwave anisotropy probe (wmap) observations: cosmological parameter
  results},  {\em The Astrophysical Journal Supplement Series} {\bf 208}
  (2013), no.~2 19.

\bibitem{xu2010cosmological}
L.~Xu and J.~Lu, {\it Cosmological constraints on generalized chaplygin gas
  model: Markov chain monte carlo approach},  {\em Journal of Cosmology and
  Astroparticle Physics} {\bf 2010} (2010), no.~03 025.

\bibitem{shi2011effects}
K.~Shi, Y.-F. Huang, and T.~Lu, {\it The effects of parametrization of the dark
  energy equation of state},  {\em Research in Astronomy and Astrophysics} {\bf
  11} (2011), no.~12 1403.

\bibitem{qi2016transient}
J.-Z. Qi, R.-J. Yang, M.-J. Zhang, and W.-B. Liu, {\it Transient acceleration
  in f (t) gravity},  {\em Research in Astronomy and Astrophysics (RAA)} {\bf
  16} (2016), no.~2 22.

\bibitem{lazkoz2012first}
R.~Lazkoz, A.~Montiel, and V.~Salzano, {\it First cosmological constraints on
  the superfluid chaplygin gas model},  {\em Physical Review D} {\bf 86}
  (2012), no.~10 103535.

\end{thebibliography}\endgroup

\end{document}